\documentclass{pic2012}
\def\runauthor{F. Arneodo}
\def\shorttitle{Dark Matter Searches}
\begin{document}

\title{Dark Matter Searches
}

\author{Francesco Arneodo}

\address{INFN - Laboratori Nazionali del Gran Sasso, and GSSI, L'Aquila, Italy}

\maketitle

\abstracts{I will review the present status of Dark Matter searches. The experimental scenario is presently very active and controversial, as some experiments do find positive signals, while others set strong limits. More data are needed, and a discovery may be very near.}

\section{Introduction: evidence of Dark Matter} 
\begin{figure}[!thb]
\begin{center}
\includegraphics[width=0.80\linewidth]{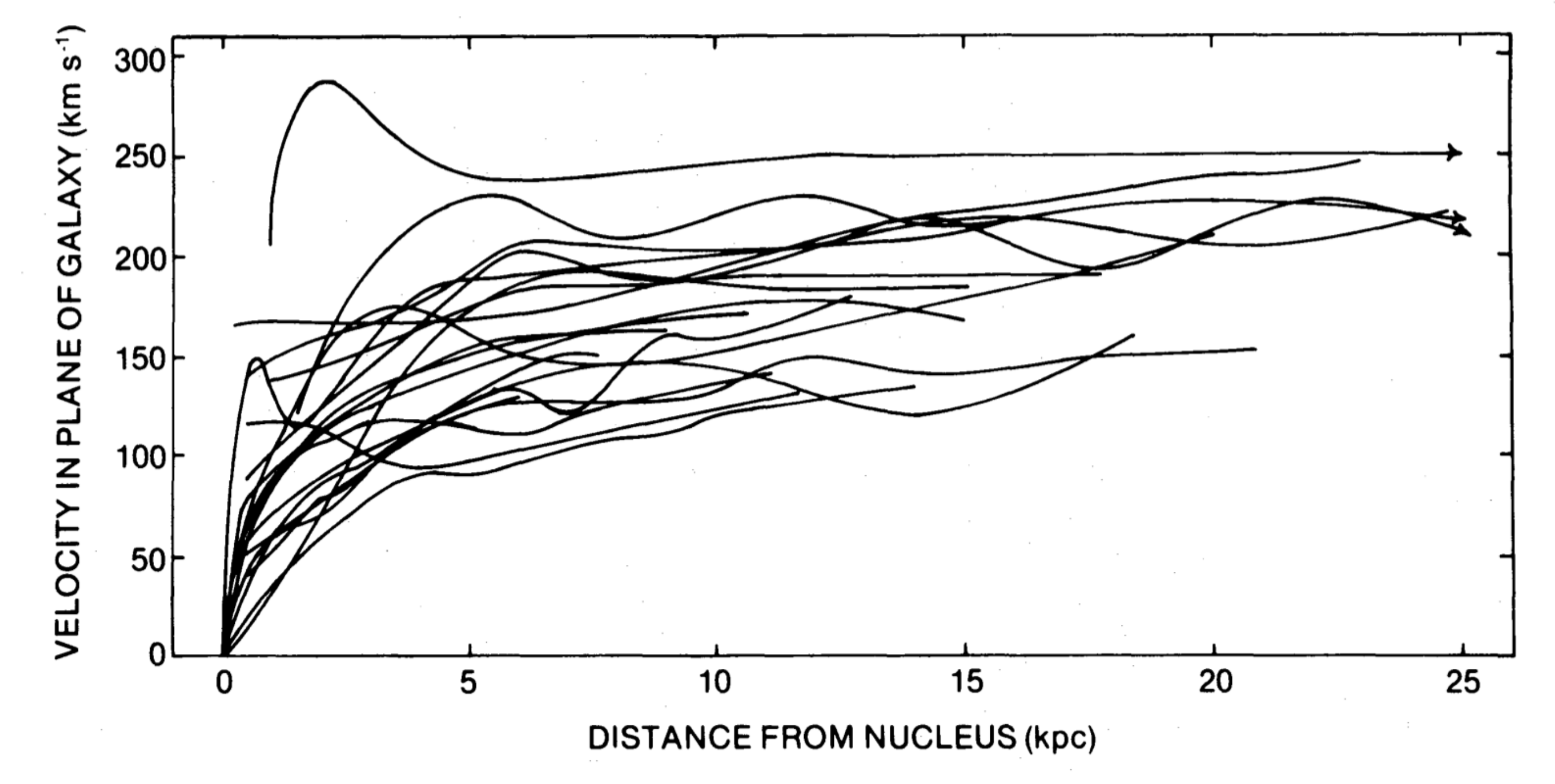}
\caption[*]{Rotational curves of 21 spiral galaxies as measured by \cite{rubin1980}}
\label{fig:rotgal}
\end{center}
\end{figure}

The existence of "dark matter" (DM in the following) was  first pointed out by F. Zwicki \cite{Zwicky:1933,Zwicky:1937p764}.  The remarkable Swiss astronomer demonstrated that  inferring the mass of a galaxy from its luminosity can provide, at best, only lower limits. Instead, by applying the  virial theorem of classical mechanics to the Coma cluster, he concluded that the galaxies in that cluster should have an average  mass of $4.5 \times 10^{10}$ solar masses, corresponding to mass/ luminosity ($\gamma$) ratio of about 500, in striking contrast with values of $\gamma=3$ considered as normal. 
Several years after, Rubin, Ford, and Thonnard \cite{rubin1980}  measured the rotational profiles of several spiral galaxies. The shape of those curves (Fig. \ref{fig:rotgal}) "implies that the mass is not centrally condensed" and that "the conclusion is inescapable that non-luminous matter exists beyond the optical galaxy" \cite{rubin1980} (p. 485).
Further evidence of the existence of Dark Matter comes from the measurements of the Cosmic Microwave Background  anisotropies, and particularly from the WMAP experiment \cite{wmap}.  According to the presently widely accepted cosmological model $\Lambda$CDM \cite{cdm} the total contribution of matter to the overall energy density is 23\%, while only 4\% is contributed by baryonic matter, this gap being filled by "cold dark matter". \\
One of the most popular proposed solutions to the Dark Matter problem is the existence of massive particles characterised by extremely weak interactions with ordinary matter, normally referred to as 'Weakly Interacting Massive Particles' (WIMPs).  A natural candidate for WIMPs is the lightest supersymmetric particle, the neutralino ($\chi$). 
Many techniques for WIMP search assume a $\approx$100 GeV neutralino with a $\chi$-nucleon cross section of  $\approx 10^{-40}$ cm$^{2}$ or less. \\
\section{Indirect searches}
In the supersymmetry framework, neutralinos are Majorana particles, meaning the possibility  of annihilation in neutralino-neutralino scattering. We should then expect  annihilation products from regions of the sky where the abundance of dark matter is presumably high. Depending on models, the annihilation may produce several kinds of Standard Model particles. To have a  detectable signal, one has to identify  annihilation messengers that can reach our detectors and that can be reasonably disentangled from the background. There are only four viable possibilities: gamma rays, neutrinos, antiprotons and positrons. I will briefly discuss four cases:
dwarf spheroidal galaxies, the galactic center,  antimatter in cosmic rays, and neutrinos from the Sun. 
\subsection{Dwarf spheroidal galaxies}
These are faint objects that are considered as satellites of the Milky Way. Some suggest that they could be the most common galaxy type in the Universe, but their faintness makes them hard to identify \cite{dwarf}. They appear to have a very high $\gamma$ (of the order of 100), the current explanation being a high DM content, making them an ideal target for observation of gamma rays from neutralino annihilation. The FERMI Large Area Telescope has recently published \cite{Ackermann:2011} a survey of the observation of 10 of these galaxies using 24 month of data. No significant excess of gamma radiation is observed. This result excludes the existence of WIMPs with annihilation cross section down to $10^{-25}$cm$^3$s$^{-1}$ for masses between 10 and 1000 GeV.
\subsection{Galactic center}
The centre of the Milky Way is another  candidate for the accumulation of Dark Matter, albeit  a  signal could be hampered by the diffuse gamma ray emission from surrounding sources. There has been  considerable attention when an analysis of FERMI-LAT public data has been published \cite{weniger2012}, showing a 130 GeV gamma bump in the spectrum from the galactic center. It might be that, at the time of the publication, other analysis have discarded this feature as a statistical or instrumental effect, however at the moment this feature is consistent with the annihilation of DM particles with a significance of 3.2 $\sigma$. More data are needed on this front, and a confirmation may come from ground-based Cherenkov arrays, such as the new  HESS-II instrument \cite{hess}.
\subsection{Anti-matter in cosmic rays}
\begin{figure}[!thb]
\begin{center}
\includegraphics[width=0.6\linewidth]{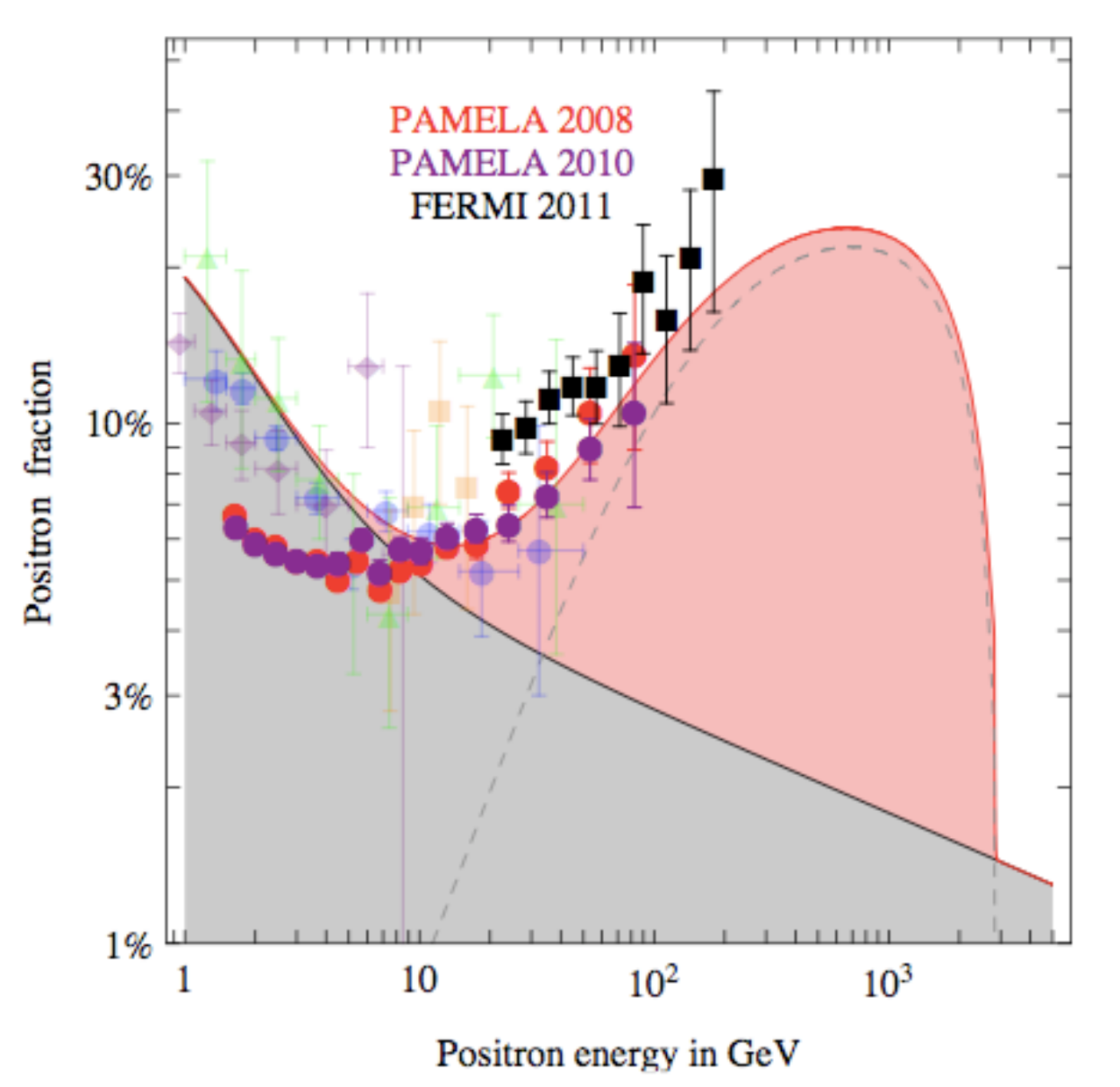}
\caption[*]{ Figure and caption by M. Cirelli \cite{cirelli}: the positron fraction of charged cosmic ray data interpreted in terms of Dark Matter annihilations: the flux from the best fit DM candidate (a 3 TeV DM particle annihilating into $\tau^+ \tau^-$ with a cross section of $2 \cdot 10^{22}$ cm$^3$s$^{-1}$) is the lower dashed line and is summed to the supposed background (solid black), giving the pink flux which fits the data.}
\label{fig:posit}
\end{center}
\end{figure}

Neutralino annihilation may give origin to an excess of antimatter in the galaxy that may be revealed locally by experiments flown on satellites, such as PAMELA \cite{pamela},  or FERMI \cite{fermi}. There are two main disadvantages in this method. First, directionality is lost, as charged particles are deflected by the galaxy's magnetic field; second, the background flux of antiprotons and positrons is not so well known. That said, both PAMELA and FERMI do see an excess of positrons in their data (see Fig. \ref{fig:posit}, from \cite{cirelli}). There are two difficulties in accepting this as a DM signal. First, no excess is seen in the antiproton flux, that would mean a kind of "leptophily" by neutralinos. Second, the fit shown in Fig. \ref{fig:posit} is compatible with  a DM particle with somewhat too high mass (3 TeV) and too high annihilation cross section.
\subsection{Neutrinos from the Sun}
 In its motion through the galaxy during the last five billions years the Sun has swept different regions of the DM halo. Given the high density of solar matter, once they start interacting with it, WIMPs start to lose energy, albeit very slowly, eventually accumulating in the Sun's core. Their annihilation should be visible by observing high energy neutrinos from the Sun, whose ordinary neutrino energy spectrum extends up to around 20 MeV. The advantage of this technique is mainly its independency from halo models,  because the Sun's motion through the Milky Way should have averaged out most halo substructures. The best apparatus to carry out this kind of observations is IceCube \cite{icecube1,icecube2}. No excess has been observed, and this translates to a very competitive limit for SD interactions (the Sun being essentially a proton target), excluding SD cross section  down to almost $ 10^{-40}$cm$^2$ for masses around 100 GeV  \cite{icecube1,icecube3}.
\section{Direct Searches}
\begin{figure}[!thb]
\begin{center}
\includegraphics[width=0.98\linewidth]{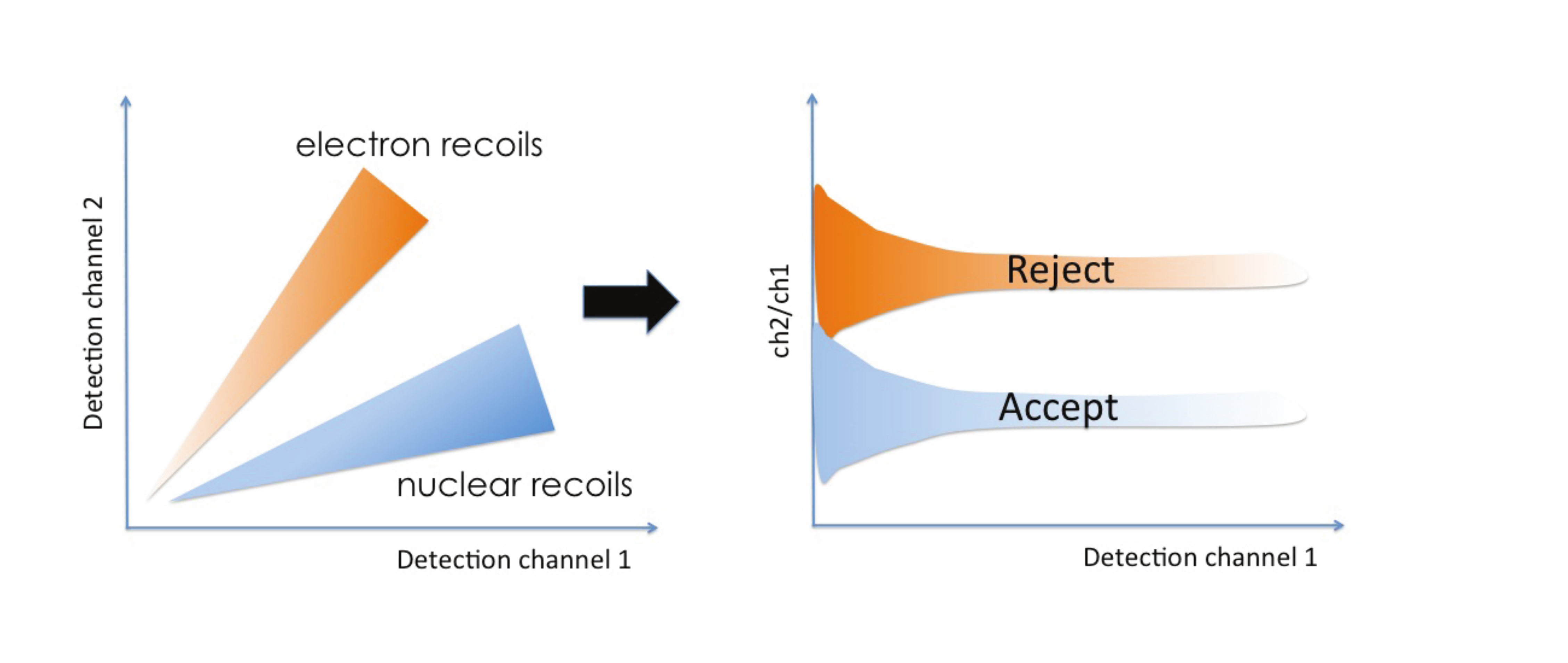}
\caption[*]{Many DM direct search experiments use two detection channels out of the three available (scintillation light, ionisation charge, phonons). Because the yield of nuclear recoils in one of the channels is usually different from electron recoils, two separated events populations are identifiable  in the Ch1 vs Ch2 phase space (left). A rejection criterion  can then be defined (right) base on pre-defined acceptance regions. More advanced analysis approaches (e.g. likelihood ratio \cite{like}) are often used. Image courtesy of E. Pantic \cite{pantic}.}
\label{fig:discr}
\end{center}
\end{figure}

Assuming a local density of the DM halo of the order of 0.3 GeV/cm$^3$,  a  candidate WIMP with 100 GeV, and a  speed of the Earth
through the  DM halo of 230 km/s, we get a local  WIMP flux of about 10$^5$cm$^{-2}$s$^{-1}$.  This would seem like a reasonable flux, if only the  cross section wouldn't be so small! The simple idea of direct detection is based on the possibility to detect the nuclear recoils originated by the rare interactions, if any, of the DM particles with the target nuclei. Two kinds of interactions can be envisaged, a scalar coupling ("spin independent " - SI), where the WIMP couples to the nucleus as a whole ($\propto A^2$), or a vector coupling ($\propto J(J+1)$) normally referred to as "spin dependent" (SD). The expected rate of interaction is:
\begin{equation}
R \approx N\frac{\rho_\chi}{m_\chi} \sigma_{\chi N} <v>
\end{equation}
where N is the number of target nuclei in the target; $\rho_\chi$ the density of DM particles; $ {m_\chi}$ the mass of the DM particle; $\sigma_{\chi N}$ is the cross section for WIMP-nucleus elastic scattering; $<v> $ is the speed of the Earth with respect to the DM halo. 
Working out $R$ for the values of $ {m_\chi}$ and  $\sigma_{\chi N}$ still allowed for supersymmetry by LHC \cite{lhc}, we get a rate of little more than one event per 1 t per year. The typical recoil energy would be of the order of 10 keV. Such a low rate of extremely low energy events would never be detectable in any above ground location because of the constant flux of charged cosmic rays. This is why all direct  Dark Matter experiments are located in underground laboratories. The underground environment takes advantage of a reduced, but definitely not zero, residual flux of high energy muons \cite{lvd}. The residual muons  and associated neutron flux, and the radioactivity of the rocks motivate the need for further shielding of the apparatuses looking for  DM interactions. The typical DM detector has an 'onion-like' shielding structure where each  layer targets a specific background. Water or polyethylene, given their hydrogen content, stop neutrons and, if the water is made active with photomultipliers, also may tag muons and associated particles. High purity copper and lead, typically placed in the inner layers, stop gamma radiation to enter the detector. Finally,  discrimination techniques, based either on the shape of the pulses or on the use of multiple detection channels (scintillation light, ionisation charge or phonons)  further enhance the signal to noise ratio, allowing to single out samples of nuclear recoil events with high efficiency. This is illustrated in Fig. \ref{fig:discr}.

\begin{figure}[!thb]
\begin{center}
\includegraphics[width=0.9\linewidth]{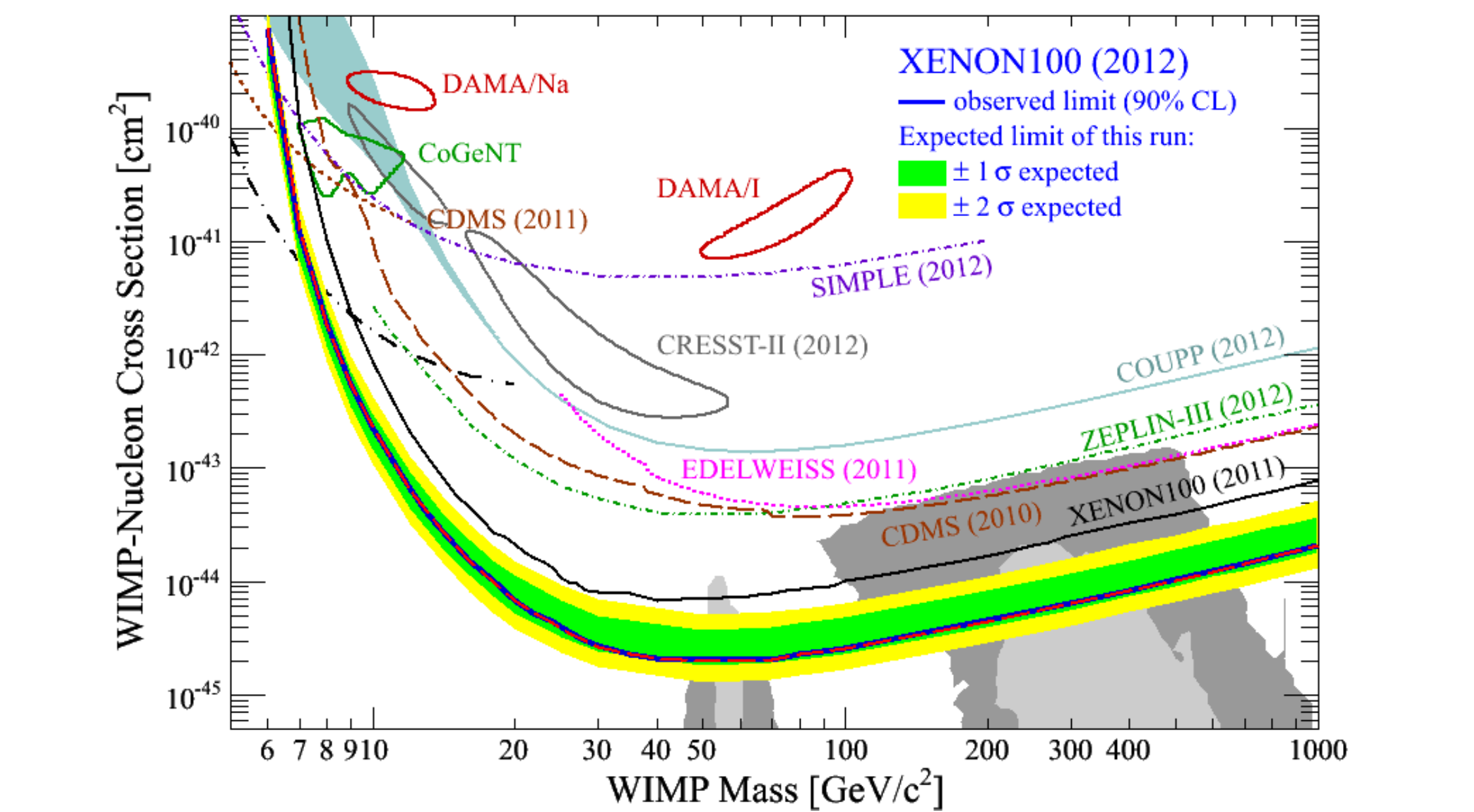}
\caption[*]{Current experimental situation of direct dark matter searches in the framework of spin-independent,  WIMP-nucleus
elastic interaction. Figure taken from \protect\cite{prl12}.}
\label{fig:scenario}
\end{center}
\end{figure}

The current experimental scenario is shown in Fig. \ref{fig:scenario}. This is the usual plot where many experiments compare their results in the space of WIMP-nucleus elastic cross section vs WIMP mass. This implies the assumption of an interaction model that may distort the true comparison of different results. In fact, in the same figure we can see positive results (see DAMA, CRESST, and COGeNT, described in the next paragraphs) along with exclusion curves from other experiments. 
It would be impossible in this short paper to give a full account of the intense experimental activity in this field. I will give some highlights, referring the interested reader to the bibliography section. I begin with the DAMA-LIBRA experiment \cite{dama1,dama2}. This one uses scintillation light only, with 250 kg of high purity NaI organised in a matrix of 25 crystals seen by  photomultipliers. The whole setup is characterised by extremely stable conditions that allow to reach a threshold as low as 2 keV.
DAMA-LIBRA observes a seasonal variation of the event rate with a maximum around end of May and one year phase. This modulation is indeed what is expected from the relative motion of the Earth with respect to the DM halo \cite{spergel}. DAMA-LIBRA's signal is now firmly established over more than 8 years. However, to accept this as a true DM signal, an independent confirmation is needed. A first approach to investigate the nature of the signal could be to repeat the experiment  elsewhere, and possibly in a different hemisphere, to rule out local effects. This is what the DM-Ice \cite{dmice} program is trying to achieve. The idea is to deploy NaI crystals  into the South Pole ice, using the same infrastructure of the IceCube experiment. The program has started with a pilot phase with two 8.5 kg crystals with the final goal of deploying a 250 kg detector. \\
Other experiments using inorganic crystals as DM targets are:  ANAIS \cite{anais}, presently taking data with a 9.6 kg NaI detector at the  Canfranc Laboratory \cite{canfranc}; and  KIMS \cite{kims}, with 100 kg of CsI operated at the relatively shallow site of Yangyang \cite{yang}, in South Korea. 

\subsection{Charge only: CoGeNT}
CoGeNT, at the Soudan Laboratory \cite{soudan}, makes  use of commercial  Ge detectors from Canberra \cite{cogent1,cogent2}. Thanks to the very low input capacitance of these crystals, they are able to achieve thresholds as low as 400 eV. In 15 months of data taking (interrupted by a fire accident in the laboratory) CoGeNT has observed an unidentified excess of events below 3 keV \cite{cogent1,cogent2} and a hint (2.8 $\sigma$) of seasonal modulation that could be compatible with DAMA. To investigate further these intriguing effects, the collaboration has started a second generation detector (called C-4 \cite{c4}) that makes use of four heavier (1.3 kg) detectors.
\subsection{Charge and phonons: CDMS at Soudan}
In the same Soudan laboratory, and a short distance from CoGeNT, we can find  another  experiment, CDMS, started in 2002 (initially in a shallow site at Stanford). The present setup, named CDMS-II, has been installed and operated since 2006 \cite{cdms1}. It is composed by 19 Ge  (250 g each) and 11 Si detectors (100 g each) cooled at $\approx 40$ mK. The detectors have a disk shape (7.6 cm diameter, 1 cm thick). The ionisation electrons  are drifted by a 3 V/cm field and collected by two concentric electrodes on one face. On the other face, four superconducting transducers collect the athermal phonons generated as well by the interaction. Combining the signals from the two channels gives a powerful way to discriminate  nuclear recoils from electron recoils. This is possible because the ionisation yield of a nuclear recoil is lower than in the case of electron recoil. 
CDMS published in 2010  the results of a total exposure of 121.3 kg-days \cite{cdms1}. The expected background in the nuclear recoil acceptance region was of $0.9\pm0.2$ events, while two events were seen, a result clearly compatible with a fluctuation of the background. In a dedicated paper  \cite{cdmsannmod}, CDMS does not see any hint of annual modulation  although their energy threshold is higher (5 keV) than CoGeNT. A new generation detector, SuperCDMS, with 100 kg mass and improved detector layout \cite{supercdms}, is being planned at SNOLAB in Canada \cite{snolab}.



\subsection{Phonons and light: CRESST at Gran Sasso}
CRESST  \cite{cresst}, located at the Gran Sasso Laboratory \cite{lngs} is another cryogenic experiment using about 10 kg of target in the form of calcium tungstate (CaWO$_4$) cylindrical crystals with a mass of 300 g each. Each crystal is equipped with two readout channels, made with Transition Edge Sensors \cite{tes}. The bottom sensor acts as a phonon detector, while the top one, thanks to a layer of silicon light absorber, detects scintillation photons. The scintillation yield of heavier particles tends to be lower, and this allows to single out nuclear recoils from the electron background. A particularly worrisome source of background is represented by surface  decays of $^{210}$Po: if the resulting alpha particle escapes the crystal, only the $^{206}$Pb may be detected, perfectly mimicking a genuine nuclear recoil. CRESST published \cite{cresst} the result of the analysis of 730 kg days. They saw 67 events in their acceptance region, with an expected background of about 44 events. If interpreted as a true signal, it would mean a WIMP of a mass around 20-50 GeV (see Fig. \ref{fig:scenario}).
Improvements of the detector setup are under way, specifically addressing some of the sources of background, such as the $^{210}$Po mentioned above. 
\subsection{Charge and light: double phase TPCs}
\begin{figure}
\centering
\includegraphics[width=0.8\linewidth]{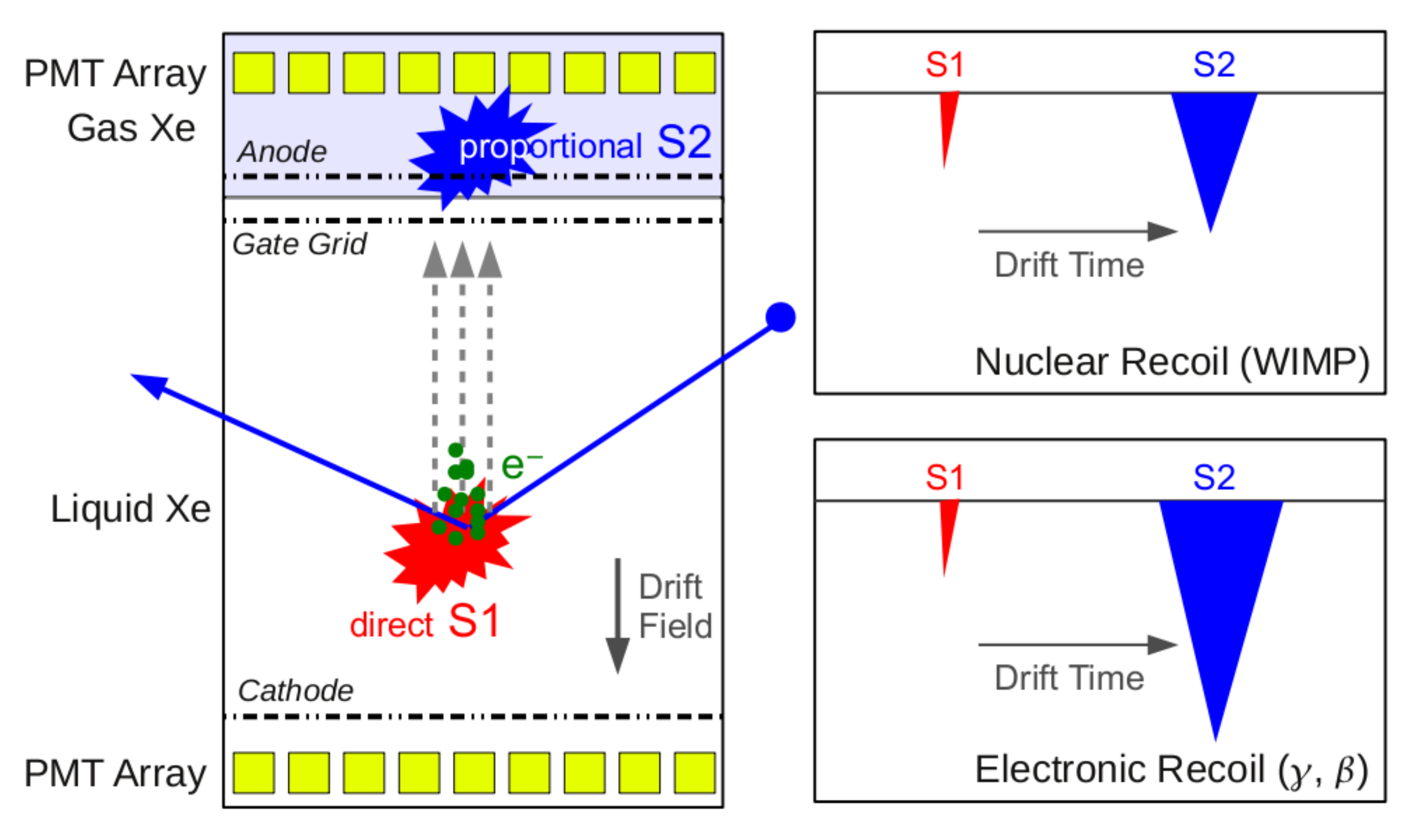}
\caption{(Left) Working principle of the XENON two-phase liquid-gas time projection
chamber (TPC). See text for details. (Right) Sketch of the waveforms of two types of
events. The different ratio of the charge (S2) and the light (S1) signal allows for the
discrimination between nuclear recoils from WIMPs and neutrons and electronic
recoils from gamma- and beta-background. Figure taken from \protect\cite{astro11}.}
\label{fig:2phase}
\end{figure}

Time projection chambers with noble liquids such as Ar and Xe are powerful and easily scalable detectors. The principle is shown in Fig. \ref{fig:2phase}. The detector's core system is a cryostat containing the noble liquid with a layer of gas at its top. Both phases are exploited by this technique, as explained below. An interaction in the liquid produces direct scintillation photons (S1) and ionization electrons. An electric field is applied across the volume with appropriate potentials on a series of electrodes, drifting ionization electrons away from the interaction zone. Electrons which reach the liquid-gas interface are extracted into the  gas, where another scintillation signal (S2), proportional to the ionisation charge, is produced.  Both the  S1 and the S2 scintillation signals are detected by photomultiplier tubes.  The ratio S2/S1 produced by a WIMP (or neutron) interaction is different from that produced by an electromagnetic interaction, allowing a rejection of the majority of the gamma and beta particle background. This detection principle has been adopted both with liquid xenon (LXe) and liquid argon (LAr). There are important differences between the two liquids that make them complementary targets. Argon has two main advantages: to be cheap, and to allow for a powerful pulse shape discrimination between nuclear and electron recoils \cite{discr}. On the other hand, the presence of the  cosmogenic isotope $^{39}$Ar  \cite{ar39} forces either to  purify it or looking for 'depleted' Ar deep in mines \cite{depar} both options making it less affordable.  Furthermore, the scintillating wavelength in LAr is about 128 nm, making it necessary to have some system to shift the light towards the sensitivity of ordinary PMTs. Xenon is much more expensive (more than 500 USD/kg), but there are no intrinsically radioactive isotopes, although small contaminations of $^{85}$Kr should be removed by distillation to allow reaching the highest sensitivities.  The scintillating wavelength is around 175 nm; its high atomic number and  natural blend of even and odd spin isotopes make it a suitable target for both SI and SD  interactions.

\begin{figure}
\centering
\includegraphics[width=0.80\linewidth]{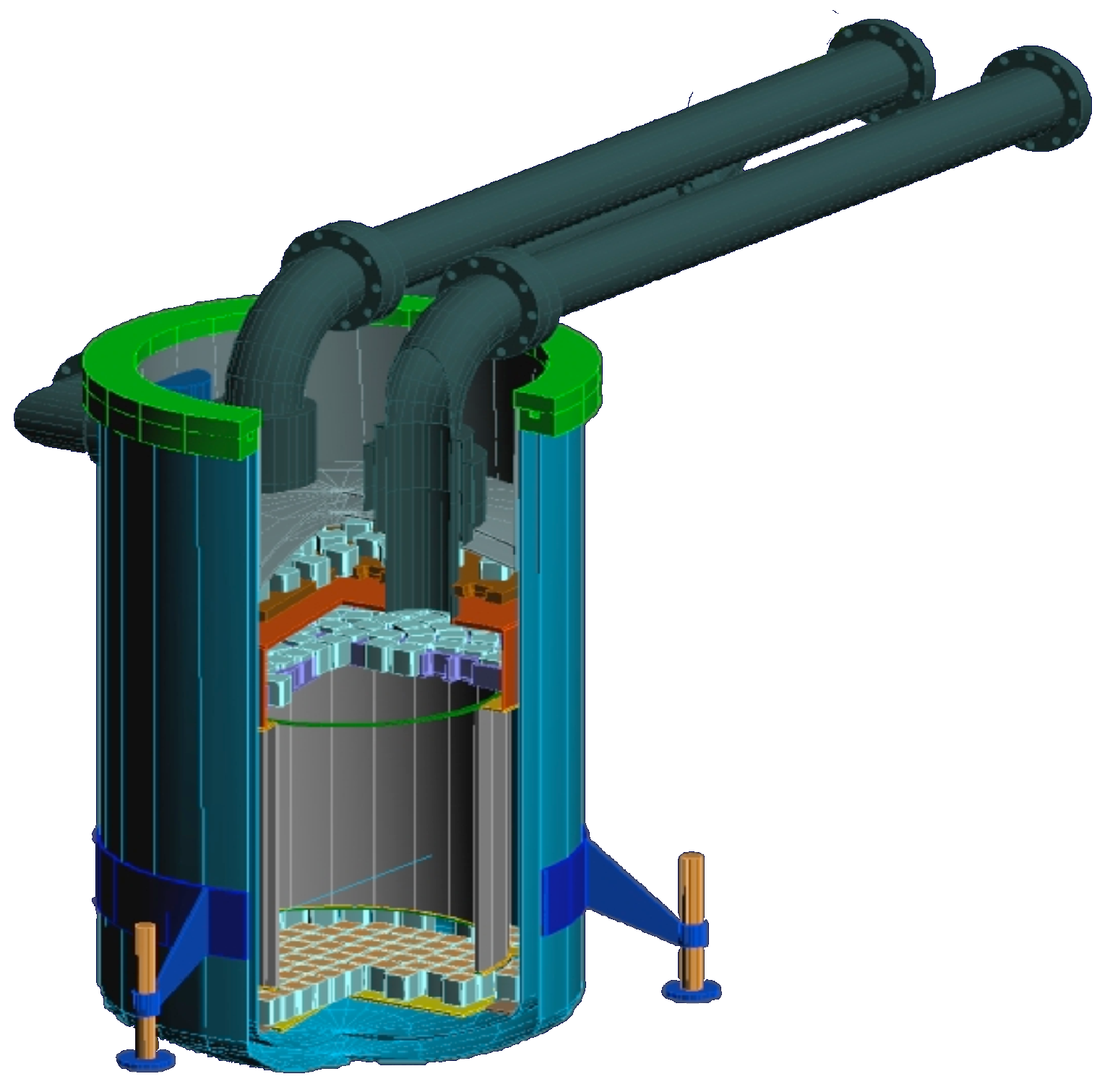}
\caption{The XENON100 dark matter detector: the inner TPC contains
62 kg of liquid xenon as target and is surrounded on all sides by an active liquid
xenon veto of 99 kg.}
\label{fig:cad}
\end{figure}

There are several projects making use of this technique, like ArDM  \cite{ardm} at Canfranc, LUX \cite{lux} at Sanford \cite{sanford}, DarkSide \cite{ds} at Gran Sasso. I will spend a few lines illustrating the XENON program, currently considered the most advanced LXe experiment. 
The first XENON detector, called XENON10, has been operated at the Gran Sasso Laboratory (LNGS) from 2006 - 2007 \cite{astro11}, achieving some of the best limits on WIMP dark matter. 
Successively, a new TPC with a factor of 10 more mass and a factor of 100 less electromagnetic background was designed to fit inside the improved passive shield of XENON10.  The present detector, XENON100 (Fig. \ref{fig:cad}), is characterised by a careful selection of all detector materials regarding intrinsic radioactivity \cite{radio}, a xenon target with lower $^{85}$Kr contamination, a novel detector design leaving only low radioactive components close to the target, and by an improved  passive shield. Furthermore, XENON100 features an active LXe veto.  The energy response of LXe at low recoil energy has been measured with a dedicated setup \cite{leff}.
XENON100 has set the most stringent limit for a very large range of WIMP masses \cite{prl11,prl12} and is currently the highest sensitivity  LXe TPC in operation. The current limit has  a minimum of 2 $\times$ 10$^{-45}$ cm$^2$ at 55 GeV and 90\% confidence level (see Fig. \ref{fig:scenario} \cite{prl12}). \\
While the XENON100 detector is still running,  the next generation detector, XENON1T, with a fiducial mass of about 1t and a total mass of 2.5t, has been designed. XENON1T will be installed in the Hall B of the Gran Sasso Laboratory, starting in 2013. 
\subsection{Alternative detectors}
I would like to finish this short review mentioning some very promising detectors. The first is COUPP \cite{coupp}. It is a bubble chamber filled with 3.5 l of CF$_3$I. It has a very powerful background rejection, because the temperature and pressure of the chamber can be tuned so to avoid any bubble nucleation by electron recoils. Furthermore, alpha decays and nuclear recoil can be discriminated thanks to the difference of the emitted sound \cite{bubbles}. Their results are very competitive for SD interactions. A bigger detector is planned for the Soudan laboratory. \\
PICASSO at SNOLAB makes use of a variant of the bubble chamber techniques, by immersing superheated droplets of C$_4$F$_{10}$ in a gel \cite{picasso}. Nuclear recoils or alpha decays will cause a droplet to explode with an acoustic signal \cite{picasso2}. The technique, as COUPP, is particularly effective for SD interactions and low mass WIMPs.
\section{Conclusions}
In this short and necessarily incomplete review I have tried to give a feeling of the present status of Dark Matter searches.  It is a highly challenging field, encompassing very different experimental techniques. Some experiments do have hints of positive signal but up to now the overall picture looks inconsistent, unless some ad hoc theoretical explanations are invoked to accommodate all results. Given the difficulty of detecting extremely rare and low energy interactions, it is very important that experiments check carefully their systematics, backgrounds, and energy scales. Given the strong efforts that are put by many excellent groups, a discovery in the next few years may be likely.
\section{Acknowledgements}
I wish to thank Prof. Dusan Bruncko and all the organising committee for having me at this Conference. I also thank Dr. Livia Ludhova, convener of my session, for a careful review of the manuscript.

\end{document}